\documentclass{article}
\usepackage{times}
\usepackage[pdfmark,colorlinks]{hyperref}
\usepackage{amsfonts}
\begin{document}
\title{Generalised Complex Geometry\\and the Planck Cone}
\author{Jos\'e M. Isidro\\
Instituto de F\'{\i}sica Corpuscular (CSIC--UVEG)\\
Apartado de Correos 22085, Valencia 46071, Spain\\
{\tt jmisidro@ific.uv.es}}

\maketitle

\begin{abstract}

Complex geometry and symplectic geometry are mirrors in string theory. The recently developed {\it generalised complex geometry}\/ interpolates between the two of them. On the other hand, the classical and quantum mechanics of a finite number of degrees of freedom are respectively described by a symplectic structure and a complex structure on classical phase space. In this letter we analyse the role played by generalised complex geometry in the classical and quantum mechanics of a finite number of degrees of freedom. We identify generalised complex geometry as an appropriate geometrical setup for dualities. The latter are interpreted as transformations connecting points in the interior of the {\it Planck cone}\/ with points in the exterior, and viceversa. The Planck cone bears some resemblance with the relativistic light--cone. However the latter cannot be traversed by physical particles, while dualities do connect the region outside the Planck cone with the region inside, and viceversa.

\end{abstract}

\tableofcontents

\section{Introduction}\label{labastidahijoputa}

Generalised complex geometry \cite{HITCHIN, GUALTIERI} turns out to have many interesting applications in supersymmetry, supergravity, 
strings and M--theory \cite{KAPUSTIN}--\cite{AS}. In this letter we elaborate on yet one more application of generalised complex structures, 
this time to the classical and quantum mechanics of a finite number of degrees of freedom: the notion of {\it duality}\/. 
This concept implies that {\it quantum}\/ vs. {\it classical}\/ is relative, or dependent on the theory one measures from \cite{VAFA}. 
That is, to what extent a given physical phenomenon is classical or quantum may be observer--dependent \cite{VAFA}. 
These ideas are largely motivated in M--theory, where there is an active interplay between physics and geometry \cite{BRUZZO}.

\section{Different geometries on phase space and their corresponding mechanics}\label{barbobketedenporkulomarikondemierda}

\subsection{Symplectic geometry: classical mechanics}\label{labastidachupamelapolla}

The clasical dynamics of a finite number $n$ of degrees of freedom is best described in terms of a classical phase space ${\cal C}$. 
The latter is at least a Poisson manifold, when not a symplectic manifold. One can use Dirac structures \cite{COURANT} 
to interpolate between Poisson structures and symplectic structures. Here we will take the symplectic point of view and assume that 
${\cal C}$ is real $2n$--dimensional admitting a symplectic structure with a symplectic form $\omega$. 
Consider the tangent and cotangent bundles to classical phase space, $T{\cal C}$ and $T^*{\cal C}$. 
The symplectic structure can be viewed as an isomorphism $\omega_x$ between the tangent and the cotangent fibres over each point 
$x\in {\cal C}$,
\begin{equation} 
\omega_x\colon T_x{\cal C}\longrightarrow T^*_x{\cal C},
\label{labastidanecesitasvaselinaparaelkulo}
\end{equation}
satisfying 
\begin{equation}
\omega_x^*=-\omega_x
\label{labastidamelasopla}
\end{equation}
for all $x\in {\cal C}$. The asterisk denotes the linear dual. The integrability condition ${\rm d}\omega=0$ must be satisfied \cite{ARNOLD}. That the cotangent bundle $T^*{\cal C}$ has a deep classical--mechanical meaning is well known \cite{ARNOLD}. This interpretation is strengthened by the use of Darboux coordinates $q^j,p_j$ around $x\in{\cal C}$, in terms of which $\omega_x$ reads
\begin{equation}
\omega_x=\sum_{j=1}^n{\rm d}q^j\wedge{\rm d}p_j.
\label{labastidaketefollenmarikon}
\end{equation}

\subsection{Complex geometry: quantum mechanics}\label{labastidakasposo}

Quantum--mechanically one is limited by Heisenberg's uncertainty principle, so the simultaneous specification of ${\rm d}q^j$ and ${\rm d}p_j$ 
in eqn. (\ref{labastidaketefollenmarikon}) has a lower bound given by $\hbar/2$. From a geometrical perspective, 
quantum mechanics abandons symplectic geometry. Instead the holomorphic tangent bundle appears naturally as the bearer of quantum--mechanical 
information about the system. Two categories arise here: holomorphic objects and tangent objects. The quantum theory requires both; let us explain why.

That the category of complex manifolds arises naturally in quantum mechanics is best appreciated in the theory of coherent states \cite{PERELOMOV}. In connection with duality transformations this point has been analysed in ref. \cite{HOL}. There is also a simple heurisitic argument in favour of holomorphic objects as appropriate for carrying quantum--mechanical information. Namely, holomorphic objects naturally respect the limitations imposed by Heisenberg's principle because, roughly speaking, they depend on $z^j=(q^j+{\rm i}p_j)/\sqrt{2}$ but not on $\bar z^j=(q^j-{\rm i}p_j)/\sqrt{2}$. In this way the transformation from Darboux coordinates $q^j, p_j$ to holomorphic coordinates $z^j$ 
cannot be inverted, since inverting it would require using also the $\bar z^j$, thus spoiling holomorphicity. In other words, the theory expressed 
in terms of holomorphic coordinates $z^j$ only contains half as much information as the theory expressed in terms of Darboux coordinates 
$q^j, p_j$, and Heisenberg's principle is respected. Equivalently we may state that the passage from classical to quantum mechanics implies a certain loss of information, 
which is implemented mathematically through complexification of classical phase space.

Quantisation, however, does not stop at complexification. Once within the holomorphic category, we will further argue in favour of holomorphic {\it tangency}\/ as being quantum--mechanical in nature. This point has been established in ref. \cite{TANGENT} for some particular examples. For the moment let us recall that a complex structure $J$ on ${\cal C}$ is an endomorphism of the tangent fibre over each point $x\in{\cal C}$
\begin{equation}
J_x\colon T_x{\cal C}\longrightarrow T_x{\cal C}
\label{labastidatehanabiertolaspataspordetras}
\end{equation}
satisfying 
\begin{equation}
J_x^2=-{\bf 1}
\label{barbonketefollen}
\end{equation}
for all $x\in {\cal C}$, as well as the Newlander--Nirenberg integrability condition that the Nijenhuis tensor $N$  vanish identically \cite{KOBAYASHI}.
The latter is defined on holomorphic tangent vectors $Z$, $W$ on ${\cal C}$ in terms of the Lie bracket $[\cdot\,,\cdot]$ of vector fields as
\begin{equation}
N(Z,W):=[Z,W]-[JZ,JW]+J[JZ,W]-J[Z,JW].
\label{labastidakabron}
\end{equation}
So the definition of a complex structure requires the notion of tangency. 

The quantum nature of tangent vectors to classical phase space follows from the previous considerations. Let the classical Darboux coordinates $q^j$, $p_j$ quantise to the quantum observables  $Q^j$, $P_j$. In the quantum theory, commutators arise as a natural composition law for operators. Commutators satisfy the same formal properties of a Lie bracket, which is the natural operation defined on tangent vectors. Hence we can think of the holomorphic tangent bundle $T_{(1,0)}{\cal C}$ as being quantum--mechanical in nature.  Above,
$T{\cal C}\otimes\mathbb{C}=T_{(1,0)}{\cal C}\oplus T_{(0,1)}{\cal C}$. By contrast, the cotangent bundle $T^*{\cal C}$ was seen to be the relevant object in classical mechanics. We conclude that the quantum theory is best expressed in terms of {\it holomorphic, tangent}\/ objects. 

A comment is in order. When $N\neq 0$ one calls $J$ an {\it almost complex structure}. Locally on ${\cal C}$, although not globally, the latter also succeeds in implementing the loss of information characteristic of the passage from classical to quantum. Now any symplectic manifold is an almost complex manifold \cite{KOBAYASHI}. Hence, at least locally, we can always develop a quantum--mechanical theory once we are given a classical mechanics: it suffices to consider the corresponding almost complex structure on classical phase space. This latter point of view has been exploited in ref. \cite{HOL}. However, in what follows we will find it more convenient to consider ${\cal C}$ a complex manifold, rather than just almost complex.

\section{Generalised complex geometry: duality}\label{labastidaeatshit}

A central idea in generalised complex geometry is that the tangent and the cotangent bundles to a manifold ${\cal C}$ 
are treated on the same footing \cite{HITCHIN, GUALTIERI}. Thus, rather than considering $T{\cal C}$ by itself or $T^*{\cal C}$ by itself, 
one considers their direct sum $T{\cal C}\oplus T^*{\cal C}$. It is then natural to suspect that generalised complex structures on 
classical phase space provide the geometry necessary to implement dualities, the latter understood as the relativity of 
the notion of {\it classical} vs. {\it quantum} \cite{VAFA}. We devote the rest of this letter to proving that this intuition is indeed correct. 
The proof essentially boils down to an identification of the different elements entering generalised complex structures over classical 
phase space, and to their proper identification in (classical and/or quantum) mechanical language. 
We follow refs. \cite{HITCHIN, GUALTIERI} closely, omitting geometrical technicalities for brevity. 
Thus, {\it e.g.}, we will illustrate our conclusions in local coordinates around a point $x\in {\cal C}$, forgetting about global issues 
that can be taken care of by the appropriate integrability conditions.

Let our mechanics have $n$ degrees of freedom. Then the total space of the bundle $T{\cal C}\oplus T^*{\cal C}$ is real $6n$--dimensional: $2n$ dimensions for the base, $4n$ for the fibre.

\subsection{The basics}\label{ramallomelasopla}

The space $T_x{\cal C}\oplus T^*_x{\cal C}$ carries the inner product 
\begin{equation}
\langle X+\xi,Y+\eta\rangle:=\frac{1}{2}\left(\xi(X)+\eta(Y)\right),
\label{linprod}
\end{equation}
where $X,Y\in T_x{\cal C}$ and $\xi, \eta\in T_x^*{\cal C}$. It has signature $(2n,2n)$. The group $SO(2n,2n)$ acts on $T_x{\cal C}\oplus T^*_x{\cal C}$. Its Lie algebra $so(2n,2n)$ decomposes as
\begin{equation}
\left(\begin{array}{cc}
A&\beta\\
B&-A^*
\end{array}\right),
\label{labastidaketefollenporkulo}
\end{equation}
where $A\in{\rm End}(T_x{\cal C})$, $A^*\in{\rm End}(T^*_x{\cal C})$ and
\begin{equation}
B:T_x{\cal C}\longrightarrow T_x^*{\cal C}, \qquad \beta:T_x^*{\cal C}\longrightarrow T_x{\cal C}
\label{ketefollenbarbon}
\end{equation}
are skew, {\it i.e.}, $B^*=-B$, $\beta^*=-\beta$. We view $B$ as a 2--form in $\Lambda^2T_x^*{\cal C}$ via $B(X)=i_XB$. Taking $A=0=\beta$ and exponentiating,
\begin{equation}
{\rm exp}\left(\begin{array}{cc}
0&0\\
B&0
\end{array}\right)=\left(\begin{array}{cc}
{\bf 1}&0\\
B&{\bf 1}
\end{array}\right),
\label{barbonazocabronazo}
\end{equation}
we obtain the orthogonal transformation
\begin{equation}
X+\xi\longrightarrow X+\xi+i_XB.
\label{barbonazohijoputazo}
\end{equation}
These transformations are important in what follows. They are called {\it B--transformations}.

A {\it generalised complex structure}\/ over ${\cal C}$, denoted ${\cal J}$, is an endomorphism of the fibre over each $x\in {\cal C}$,
\begin{equation}
{\cal J}_x\colon T_x{\cal C}\oplus T^*_x{\cal C}\longrightarrow T_x{\cal C}\oplus T_x^*{\cal C},
\label{labastidatienesunagujeroprofundoenelkulo}
\end{equation}
satisfying the following conditions.  For all $x\in {\cal C}$ one has 
\begin{equation}
{\cal J}_x^2=-{\bf 1}
\label{labastidaketehostien}
\end{equation}
and 
\begin{equation}
{\cal J}_x^*=-{\cal J}_x.
\label{barboncortatelospelosketienescaspa}
\end{equation}
Moreover the Courant integrability condition must hold; in what follows we will assume that this condition is always satisfied. It should be realised that generalised complex geometry involves an object ${\cal J}$ that is simultaneously complex (eqn. (\ref{labastidaketehostien})) and symplectic (eqn. (\ref{barboncortatelospelosketienescaspa})). 

Suppose that ${\cal J}$ at $x\in {\cal C}$ is given by
\begin{equation}
{\cal J}_{\omega_x}=\left(\begin{array}{cc}
0&-\omega_x^{-1}\\
\omega_x&0\end{array}\right),
\label{ramallocomprateunchampuanticaspa}
\end{equation}
$\omega$ being a symplectic form as in section \ref{labastidachupamelapolla}. This ${\cal J}_{\omega}$ defines a generalised complex structure 
{\it of symplectic type}\/ ($k=0$). Indeed, ${\cal J}_{\omega}$ defines a symplectic structure on ${\cal C}$ in the sense of section 
\ref{labastidachupamelapolla}. In physical terms, this ${\cal J}_{\omega}$ describes a classical mechanics. 

At the other end we have that
\begin{equation}
{\cal J}_{J_x}=\left(\begin{array}{cc}
-J_x&0\\
0&J^*_x\end{array}\right),
\label{ramallotieneslospelosllenosdecaspa}
\end{equation}
$J$ being a complex structure as in section \ref{labastidakasposo}, defines a generalised complex structure {\it of complex type}\/ $(k=n)$. 
Again, ${\cal J}_{J}$ defines a complex structure on ${\cal C}$ in the sense of section \ref{labastidakasposo}. In physical terms, 
this ${\cal J}_{J}$ describes a quantum mechanics.

\subsection{A Darboux--like theorem}\label{ramalloguarrolavatekehuelesamierda}

There exists a Darboux--like theorem describing the local form of a generalised complex structure in the neighbourhood of any regular point. Roughly speaking, any manifold endowed with a generalised complex structure splits {\it locally}\/ as the product of a complex manifold times a symplectic manifold. A more precise statement is as follows. A point $x\in{\cal C}$ is said {\it regular}\/ if the Poisson structure $\omega^{-1}$ has constant rank in a neighbourhood of $x$. Then any regular point in a generalised complex manifold has 
a neighbourhood which is equivalent, via a diffeomorphism and a $B$--transformation, to the product of an open set in $\mathbb{C}^k$ with an open set in $\mathbb{R}^{2n-2k}$, the latter endowed with its standard symplectic form. The nonnegative integer $k$ is called the {\it type}\/ of ${\cal J}$, $k=0$ and $k=n$ being the limiting cases examined above.

\subsection{$B$--transformations}\label{ramallolimpiateelkuloguarro}

Next assume that ${\cal C}$ is a linear space. Then any generalised complex structure of type $k=0$ is the $B$--transform of a symplectic structure. This means that any generalised complex structure of type $k=0$ can be written as
$$
{\rm e}^{-B}{\cal J}_{\omega}{\rm e}^{B}=
\left(\begin{array}{cc}
{\bf 1}&0\\
-B&{\bf 1}\end{array}\right)
\left(\begin{array}{cc}
0&-\omega^{-1}\\
\omega &0\end{array}\right)
\left(\begin{array}{cc}
{\bf 1}&0\\
B&1\end{array}\right)
$$
\begin{equation}
=\left(\begin{array}{cc}
-\omega^{-1}B & -\omega^{-1}\\
\omega + B\omega^{-1}B & B\omega^{-1}\end{array}\right),
\label{ramalloestasllenodecaspa}
\end{equation}
for a certain 2--form $B$. Similarly any generalised complex structure of type $k=n$ over a linear manifold ${\cal C}$ is the $B$--transform of a complex structure,
$$
{\rm e}^{-B}{\cal J}_{J}{\rm e}^{B}=
\left(\begin{array}{cc}
{\bf 1}&0\\
-B&{\bf 1}\end{array}\right)
\left(\begin{array}{cc}
-J&0\\
0 & J^*\end{array}\right)
\left(\begin{array}{cc}
{\bf 1}&0\\
B&{\bf 1}\end{array}\right)
$$
\begin{equation}
=\left(\begin{array}{cc}
-J & 0\\
BJ+J^*B & J^*\end{array}\right).
\label{ramalloduchatekehuelesamierda}
\end{equation}
When ${\cal C}$ is an arbitrary smooth manifold, not necessarily a linear space, the previous statements hold essentially true, with some refinements required; see refs. \cite{HITCHIN, GUALTIERI}.

\subsection{The square of the symplectic form}\label{barbonhijoperra}

A metric of indefinite signature $(2n,2n)$ is readily manufactured with the mechanical elements at hand. Starting from the classical symplectic form $\omega$ in Darboux coordinates, its (block) matrix at $x\in{\cal C}$ is
\begin{equation}
\omega_x=\left(\begin{array}{cc}
0&-{\bf 1}\\
{\bf 1}&0\end{array}\right).
\label{barbonketehostienporladron}
\end{equation}
Its inverse $\omega^{-1}_x=-\omega_x$ is the matrix representing classical Poisson brackets in Darboux coordinates \cite{ARNOLD}, and $(-\omega_x)^2=-{\bf 1}$. Now ${\rm i}\hbar$ times classical Poisson brackets are quantum commutators. Hence the latter are represented by the matrix
\begin{equation}
{\rm i}\hbar\cdot\left(\begin{array}{cc}
0&{\bf 1}\\
-{\bf 1}&0\end{array}\right).
\label{barbonketehostienpormarikon}
\end{equation}
Setting $\hbar=1$, the above squares to the identity. The direct sum of the squares of the matrices (\ref{barbonketehostienpormarikon}), (\ref{barbonketehostienporladron}) gives us the expression, in local coordinates, of a diagonal metric on $T{\cal C}\oplus T^*{\cal C}$
\begin{equation}
\left(\begin{array}{cc}
{\bf 1}_{2n}&0\\
0&-{\bf 1}_{2n}\end{array}\right)
\label{barboneatshit}
\end{equation}
with the desired signature $(2n,2n)$. The symplectic structure on classical phase space, plus the quantisation prescription that ${\rm i}\hbar$ times Poisson brackets become quantum commutators, automatically dictate that the negative--signature piece of the metric must be classical, while the piece with positive signature must be quantum.

\subsection{The Planck cone}\label{elputoconyoderamallo}

Many textbooks on special relativity take the constancy of the speed of light as their starting point.  Mathematically this can be recast as the invariance of the light--cone under Lorentz transformations. The light--cone separates physical particles from tachyons, the cone itself corresponding to massless particles. The Lorentz group $SO(1,3)$ arises naturally in this setup. In our context we have the group $SO(2n,2n)$. Putting aside the fact that $Q^j$ and $P_j$ are actually quantum--mechanical operators on Hilbert space, let us temporarily treat them like c--numbers ({\it i.e.}, let us regard them as a basis of generators for the tangent space $T_x{\cal C}$). Correspondingly we can consider the cone defined within $T_x{\cal C}\oplus T_x^*{\cal C}=\mathbb{R}^{2n}\oplus\mathbb{R}^{2n}$ by the metric (\ref{barboneatshit}).
\begin{equation}
\sum_{j=1}^n\left((Q^j)^2+(P_j)^2\right)-\sum_{j=1}^n\left((q^j)^2+(p_j)^2\right)=0.
\label{labastidaketedenporelconyo}
\end{equation}
This cone separates $T_x{\cal C}\oplus T_x^*{\cal C}$ into two regions, the interior and the exterior. It  makes sense to call (\ref{labastidaketedenporelconyo}) {\it the Planck cone}.

Treating the classical realm as separate from the quantum world means that the two regions of $T_x{\cal C}\oplus T^*_x{\cal C}$ separated by the Planck cone (\ref{labastidaketedenporelconyo}) are disconnected, {\it i.e.}, no transformations are allowed between the two of them. There exists an analogous situation in relativistic physics: spacelike points are disconnected from timelike points because of a physical principle preventing the crossing of the light--cone. Namely, all physical signals must propagate at a speed $v\leq c$. What the notion of duality means is that, {\bf under certain circumstances} to be specified below, {\it classical} vs. {\it quantum} are not completely separate worlds, and transformations between them are allowed. An $SO(2n,2n)$--rotation can map points inside the cone (\ref{labastidaketedenporelconyo}) into points outside and viceversa, and a duality can be viewed as {\it a crossing of the Planck cone}. In mechanical problems that do {\it not}\/ exhibit dualities (such are all the usual examples known in the textbooks) there is an analogue of the physical principle prohibiting the crossing of the light--cone. Namely, the very inexistence of dualities itself serves as such a principle. On the other hand, there is ample evidence from string duality and M--theory \cite{VAFA} that a formulation is required for quantum mechanics, where dualities may be manifestly implemented. In fact such a formulation has been explicitly demanded in ref. \cite{VAFA}, section 6.

Let us clarify the circumstances (alluded to above in boldface) under which a duality may exist. Duality--free mechanics admit generalised complex structures of extremal types, either $k=0$ or $k=n$. Hence {\it duality transformations may exist only if the generalised complex structure is of nonextremal type $0\neq k \neq n$}. Even if $k$ remains a constant across ${\cal C}$, its being neither 0 nor $n$ means that life under the corresponding mechanics is neither fully classical nor fully quantum, as one can traverse the Planck cone back and forth by means of an $SO(2n,2n)$--rotation. In such a case the corresponding classical phase space ${\cal C}$ cannot be a symplectic manifold. At most it can be a Poisson manifold (for a review see ref. \cite{WEINSTEIN}). Examples of these manifolds can be found in refs. \cite{HITCHIN, GUALTIERI}, where the so--called {\it jumping phenomenon}\/ (the nonconstancy of $k$ across ${\cal C}$) is also illustrated. Since a generalised complex structure implies a reduction in the structure group from $SO(2n,2n)$ to $SU(n,n)$ \cite{HITCHIN, GUALTIERI}, we will henceforth refer to dualities as $SU(n,n)$--rotations.

\section{Discussion}\label{labastidatelahanmetidoporkulo}

There are deep connections between the symplectic category and the holomorphic category (for a review see ref. \cite{ORLOV}).
In brief, generalised complex geometry interpolates between symplectic geometry and complex geometry, the latter two appearing as limiting cases of one single geometry. At the classical end we have the symplectic structure; at the quantum end we have the complex structure. The novelty of generalised complex geometry lies in the fact that both structures arise as different aspects of one and the same entity, namely, a generalised complex structure. 

We have applied generalised complex geometry to the classical and quantum mechanics of a finite number of degrees of freedom. We have interpreted the cotangent bundle $T^*{\cal C}$ as {\it classical}\/ and the tangent bundle $T{\cal C}$ as {\it quantum}\/ in order to provide a natural construction of a metric on $T{\cal C}\oplus T^*{\cal C}$ carrying the required signature $(2n,2n)$. 

Nonextremal values of the type $0\neq k\neq n$ of the generalised complex structure allow for duality transformations. The latter are $SU(n,n)$--rotations acting on the fibres of the bundle $T{\cal C}\oplus T^*{\cal C}$. These rotations are always well defined on any generalised complex manifold ${\cal C}$ whatever the value of the type $k$, but they cannot be physically realised when $k=0$ or $k=n$. The existence of the $B$--field is an immediate consequence of this $SU(n,n)$ symmetry. In strings and M--theory the $B$--field is identified with the Neveu--Schwarz antisymmetric tensor. In our mechanical setup the $B$--field arises as the generator of certain $SU(n,n)$--rotations. The $B$--field also allows one to reach (noncanonically) any generalised complex structure of a given type $k$ from certain distinguished structures carrying the same type $k$. In a regular neighbourhood, a generalised complex structure gives rise to a foliation with symplectic leaves and a transverse complex structure. It is interesting to observe that this is exactly the foliation described in the 2nd ref. of \cite{HOL}, section 5.3.

The Planck cone is a pictorial way of illustrating our goal, that classical and quantum mechanics be treated on the same mathematical footing, and that duality transformations between the classical and quantum realms be allowed. In geometrical terms this is reflected in the fact that generalised complex structures refer to $T{\cal C}\oplus T^*{\cal C}$, rather than $T{\cal C}$ or $T^*{\cal C}$ alone. 

It would be interesting to investigate what applications generalised complex geometry may have on the quantum theory of gravity. In our simple, mechanical setup we have not quantised gravity, but we certainly have rendered the notion of a quantum {\it relative}.

{\bf Acknowledgements}

It is a great pleasure to thank J. de Azc\'arraga for encouragement and support, S. Theisen for conversations, and Max-Planck-Institut f\"ur Gravitationsphysik, Albert-Einstein-Institut (Potsdam, Germany) for hospitality. This work has been partially supported by research grant BFM2002--03681 from Ministerio de Ciencia y Tecnolog\'{\i}a, by research grant GV2004-B-226 from Generalitat Valenciana, by EU FEDER funds, by Fundaci\'on Marina Bueno and by Deutsche Forschungsgemeinschaft.

\end{document}